\documentclass[journal]{IEEEtran}
\usepackage{amsmath,amsfonts}
\usepackage{algorithmic}
\usepackage{algorithm}
\usepackage{array}
\usepackage[caption=false]{subfig}
\usepackage{textcomp}
\usepackage{stfloats}
\usepackage{url}
\usepackage{verbatim}
\usepackage{graphicx}
\usepackage{cite}
\usepackage{bm}
\usepackage{booktabs}
\usepackage{hyperref}
\usepackage{multirow, multicol}
\usepackage{makecell}
\usepackage{amssymb}
\usepackage{bbding}
\usepackage{enumitem}
\usepackage{arydshln}

\begin{document}
%


\title{Towards Real-world Environment-aware Zero-shot Text-to-speech Synthesis via Disentangled Audio Infilling}
%
%
%

\author{Ye-Xin Lu, 
        Xin Wang,~\IEEEmembership{Member,~IEEE,} 
        Yang Ai,~\IEEEmembership{Member,~IEEE,} 
        Hui-Peng Du,~\IEEEmembership{Student Member,~IEEE,} 
        Zhen-Hua Ling,~\IEEEmembership{Senior Member,~IEEE,}
        and Junichi Yamagishi,~\IEEEmembership{Senior Member,~IEEE}%
\thanks{This work is the extended version of our conference paper \cite{lu2026daien} published at IEEE International Conference on Acoustics, Speech and Signal Processing 2026 (ICASSP 2026).}
\thanks{Y.-X. Lu, Y. Ai, H.-P. Du and Z.-H. Ling are with the National Engineering Research Center of Speech and Language Information Processing, University of Science and Technology of China, Hefei, China (e-mail: yxlu0102@mail.ustc.edu.cn, yangai@ustc.edu.cn, redmist@mail.ustc.edu.cn, zhling@ustc.edu.cn).
X. Wang and J. Yamagishi are with the National Institute of Informatics, Tokyo, Japan (e-mail: wangxin@nii.ac.jp, jyamagis@nii.ac.jp).}
        }
\markboth{Journal of \LaTeX\ Class Files,~Vol.~14, No.~8, August~2015}%
{Shell \MakeLowercase{\textit{et al.}}: Bare Demo of IEEEtran.cls for IEEE Journals}
%



\maketitle

\begin{abstract}

Recent zero-shot text-to-speech (TTS) systems achieve remarkable naturalness and speaker similarity but typically require high-quality speaker prompts and either strip away or entangle the acoustic environment with speaker characteristics, limiting their real-world applicability. 
We present an extended DAIEN-TTS, an environment-aware zero-shot TTS framework that disentangles and jointly models speech, background noise, and reverberation, enabling independent control over timbre and acoustic environment through separate speaker and environment prompts. 
Built upon the flow-matching-based F5-TTS, it uses a speech-environment separation module to decompose environmental speech into speech, noise, and reverberation components, which are injected into the Diffusion Transformer for environment-aware generation. 
Training uses simulated data constructed by mixing clean speech with noise and room impulse responses, together with a cross-speaker conditioning strategy that suppresses speaker information leakage from the environment branch. 
When real-world data are available, the system can be further fine-tuned to bridge the simulated-to-real domain gap.
At inference, a triple classifier-free guidance mechanism enables fine-grained control over speech, noise, and reverberation, and a signal-to-noise-ratio adaptation strategy aligns the synthesized speech with the environment prompt. 
Experiments on simulated and real-world test sets show that DAIEN-TTS generates environmental personalized speech with high naturalness, strong speaker similarity, and faithful noise and reverberation reproduction, while offering controllability beyond prior environment-aware TTS systems.

\end{abstract}


%
\IEEEpeerreviewmaketitle

\section{Introduction}
%
%
%
%
\IEEEPARstart{Z}{ero-shot} text-to-speech (TTS) is used to synthesize speech that preserves the voice characteristics of a target speaker, using only a few seconds of audio prompt, without requiring any speaker-specific training data.
Current zero-shot TTS methods can be broadly categorized into two types.
Speaker-embedding-based methods \cite{jia2018transfer, cooper2020zero, casanova2022yourtts} extract a fixed speaker representation from the audio prompt which is then used as a conditioning input to control the timbre of the synthesized speech.
More recently, in-context-learning (ICL)-based methods, including neural-codec-based language modeling \cite{chen2025neural, wang2024speechx}, flow-matching-based speech infilling \cite{le2023voicebox, chen2025f5}, and their hybrids \cite{anastassiou2024seed, du2024cosyvoice, leeditto}, directly leverage the audio prompt as context for generation, achieving substantial improvements in naturalness and speaker similarity. 
Despite their differences, both types are typically trained on large-scale high-quality speech corpora, thus requiring high-quality audio prompts and tending to produce clean personalized speech.

In practice, however, audio prompts are often recorded under diverse real-world acoustic conditions, where various environmental factors such as background noise and room reverberation are inevitably present.
The two types of TTS methods exhibit different behaviors when encountering such environmental factors.
Speaker-embedding-based methods encode the speaker characteristics of the audio prompt into a global embedding using pretrained environment-robust speaker encoders \cite{snyder2018x, desplanques2020ecapa}, thus inherently discarding frame-level environmental information and producing clean synthesized speech regardless of the acoustic condition of the audio prompt.
ICL-based methods, on the other hand, tend to preserve the environmental characteristics of the audio prompt in the synthesized speech.
However, the environmental information is inherently entangled with the speaker characteristics, making it difficult to, for example, reproduce these characteristics from one recording while applying the acoustic environment from another.
When the audio prompt contains severe environmental distortions, the synthesized speech with these methods often exhibit degraded intelligibility and naturalness \cite{wang2024investigation, lu2025improving}.

In summary, current zero-shot TTS methods either discard or entangle the acoustic environment with speaker characteristics, without providing explicit control over the environment of the synthesized speech.
In many real-world application scenarios, however, the acoustic environment present in the speech signal is not merely an undesired distortion, but rather constitutes a desired auditory attribute of the synthesized speech.
For instance, in AI audiobook production, speech often needs to be rendered with background sounds that match the narrative scene, such as a bustling street or quiet room \cite{lee2024voiceldm, jung2025voicedit}.
In the generation of synthetic training data for automatic speech recognition (ASR) \cite{ko2017study, hilmes2024effect} or automatic speaker verification (ASV), realistic reproduction of diverse acoustic environments is essential to ensure that the trained models generalize well to real-world conditions.
These applications call for an environment-aware TTS system that can independently specify the speaker identity and acoustic environment of the synthesized speech.

Several studies have explored environment-aware TTS to independently control the timbre and acoustic environment of the synthesized speech through separate speaker and environment audio prompts.
Current TTS systems \cite{tan2022environment, lu2025incremental} use dedicated speaker and environment encoders to extract global embeddings for disentangled control but are limited to time-invariant acoustic environments, such as reverberation, and cannot handle time-varying background noise.
In addition, controllable masked speech prediction \cite{zhang2025advanced} has been introduced to selectively remove or preserve the background sound of the speaker prompt; however, the acoustic environment of the synthesized
speech is limited to a binary choice between preservation and removal, and cannot be specified independently of the speaker prompt.
From the text-to-audio (TTA) generation perspective, systems such as VoiceLDM \cite{lee2024voiceldm} and VoiceDiT \cite{jung2025voicedit} have the ability to generate environmental audio containing speech but are not designed for TTS and lack zero-shot speaker cloning capability or exhibit poor alignment between speech and environment.
In our prior work \cite{lu2026daien}, we proposed the TTS system DAIEN-TTS, which introduces a speech-environment separation (SES) module to disentangle the background noise from the speech prompt and achieves time-varying ENvironment-aware TTS via Disentangled Audio Infilling.
However, we only considered background noise without modeling reverberation and trained exclusively on simulated data, leaving a domain gap when applied to real-world recordings.

For this work, we extended DAIEN-TTS to address these limitations towards real-world environment-aware zero-shot TTS. Specifically, we extended the SES module to jointly model background noise and reverberation, where the noise is represented as a frame-level mel-spectrogram and the reverberation is encoded as an utterance-level embedding. 
We further introduce a cross-speaker conditioning strategy during training to prevent speaker information leakage from the environment branch. 
At inference, we extend the dual classifier-free guidance (DCFG) to a triple CFC (TCFG) mechanism, enabling independent control over the speech, noise, and reverberation components, along with a signal-to-noise ratio (SNR) adaptation strategy to align the synthesized speech with the environment prompt. 
Finally, to bridge the simulated-to-real domain gap, the system is first trained on simulated data then fine-tuned on real-world speech data. Extensive experiments on both simulated and real-world test sets demonstrate that the extended DAIEN-TTS generates environmental personalized speech with high naturalness, strong speaker similarity, faithful environment reconstruction, and generalizes well to real-world acoustic conditions.

The main contributions of this paper are summarized as follows:
\begin{enumerate}
\item We present an extended DAIEN-TTS framework that jointly models speech, background noise, and reverberation, enabling independent control over timbre and acoustic environment through separate audio prompts.
\item We present our designed two-stage SES module and cross-attention-based environment-conditioning mechanism to disentangle and reintegrate noise and reverberation into the generation process.
\item We introduce a TCFG mechanism and SNR adaptation strategy for fine-grained and independent control over the speech, noise, and reverberation components at inference.
\item We demonstrate that the extended DAIEN-TTS system trained on simulated data can be further fine-tuned on real-world speech data when available, effectively bridging the simulated-to-real domain gap.
\item We conducted extensive evaluations on both simulated and real-world test sets, demonstrating the effectiveness of the extended DAIEN-TTS framework.
\end{enumerate}



\section{Problem Formulation}

\begin{figure}[t]
  \centering
  \includegraphics[width=0.9\columnwidth]{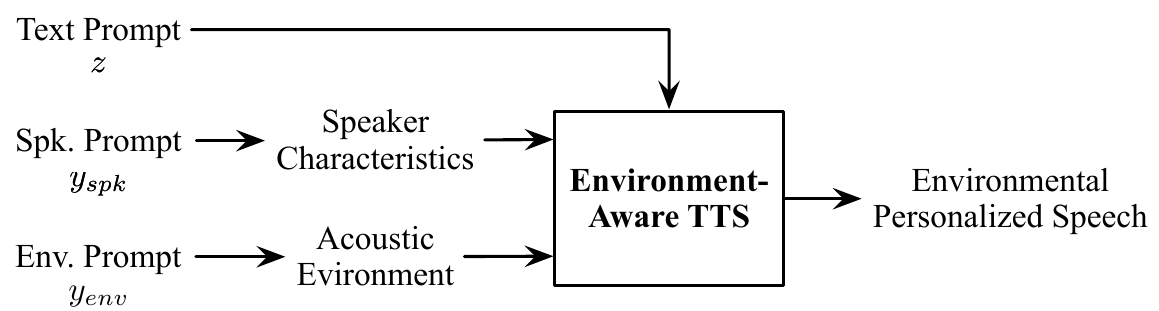}
  \caption{Task formulation of environment-aware zero-shot TTS with separate text, speaker, and environment prompts}
  \label{fig:task}
\end{figure}

In real-world speech recordings, the captured signal is not only determined by the speech content and speaker's voice but also shaped by the acoustic environment. 
The acoustic environment encompasses a variety of factors that can be broadly categorized into two types: time-varying factors, such as background noise and ambient sound effects, and time-invariant factors, such as room reverberation and channel distortions. 
Considering a clean speech signal $\mathbf{s}$, additive time-varying environmental signal $\mathbf{n}$, and room impulse response (RIR) $\mathbf{h}$ representing the time-invariant component, the recorded environmental speech can be approximated as
\begin{equation}
\mathbf{y} = \mathbf{s} * \mathbf{h} + \mathbf{n},
\end{equation}
where $*$ denotes convolution.
Throughout this paper, we use the term \textit{environmental speech} to refer to speech recorded under such non-clean acoustic conditions, where various environmental factors may be present.

Speech signals therefore carry at least three types of information: semantic content, speaker characteristics, and acoustic environment. 
The goal with environment-aware TTS is to disentangle these factors and recombine them in a controllable manner such that each can be independently specified through separate inputs. 
As illustrated in Fig.~\ref{fig:task}, given a text input $\mathbf{z}$, speaker audio prompt $\mathbf{y}_{spk}$, and environment audio prompt $\mathbf{y}_{env}$, the system synthesizes a new environmental speech that conveys the semantic content specified by $\mathbf{z}$, preserves the speaker characteristics extracted from $\mathbf{y}_{spk}$, and reproduces the acoustic environment extracted from $\mathbf{y}_{env}$. 
Both $\mathbf{y}_{spk}$ and $\mathbf{y}_{env}$ are environmental speech recordings, as expressed with Eq.~(1), from which the speaker information (i.e., encoded in $\mathbf{s}$) and environment information (i.e., encoded in $\mathbf{n}$ and $\mathbf{h}$) are respectively disentangled and recombined for generation. 
Note that we do not know ground-truth $\mathbf{n}$ and $\mathbf{h}$ for generation. 
While the text input and speaker prompt provide direct control over the semantic content and speaker characteristics using established zero-shot TTS methods, the primary challenge lies in effectively modeling and controlling the acoustic environment, which is inherently complex and composed of multiple heterogeneous factors.

On the basis of the distinct physical properties expressed with Eq.~(1), we decompose the acoustic environment into a time-varying component and time-invariant component and model them with different representations; the time-varying component is captured by a frame-level representation to preserve its temporal dynamics, while the time-invariant component is encoded as an utterance-level embedding to reflect its global characteristics. 
The details of how these representations are obtained and integrated into our TTS framework are described in the following section.

\begin{figure*}[t!]
  \centering
  \includegraphics[width=\textwidth]{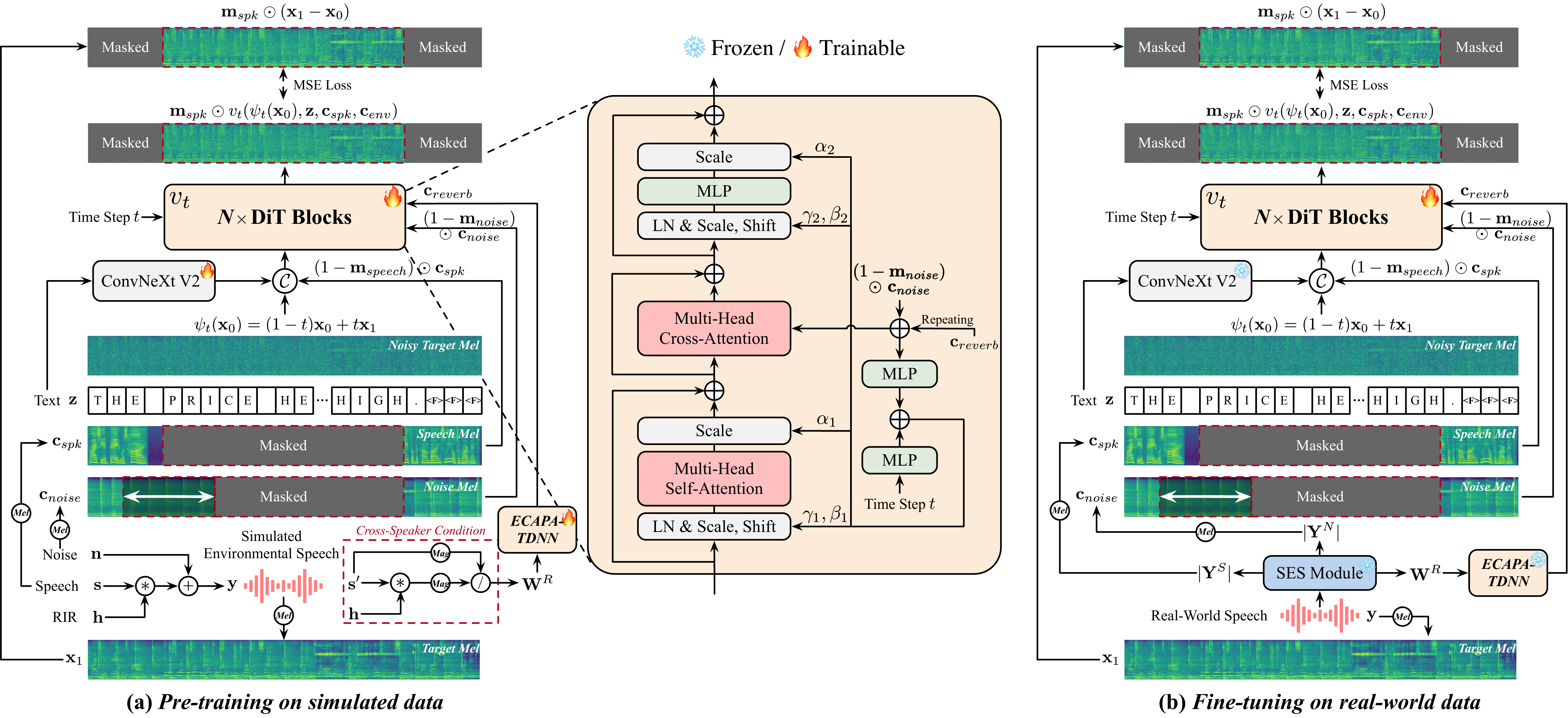}
  \caption{Overview of the two-stage training of the extended DAIEN-TTS, 
  with the detailed DiT block structure shown in the middle. 
  (a) \textbf{Pretraining on simulated data:} the noisy target $\mathbf{x}_1$ is 
  constructed from the clean speech $\mathbf{s}$ of the target speaker, 
  a selected noise $\mathbf{n}$, and RIR $\mathbf{h}$, while the 
  reverberation condition is extracted from the speech $\mathbf{s}'$ of 
  a different speaker convolved with the same $\mathbf{h}$. 
  (b) \textbf{Fine-tuning on real-world data:} neither $\mathbf{n}$ nor $\mathbf{h}$ 
  is available; instead, the SES module separates the speech and noise 
  components from real environmental recordings to provide the conditions. 
  Snowflake and flame icons denote frozen and trainable modules, respectively.}
  \label{fig: model_train}
\end{figure*}

\section{Proposed Method}
\subsection{System Overview}

The architecture of the extended DAIEN-TTS is illustrated in Figs.~\ref{fig: model_train} and~\ref{fig: ses}. 
Built upon F5-TTS \cite{chen2025f5}, a flow-matching-based zero-shot TTS framework that operates in the mel-spectrogram domain via masked audio infilling \cite{le2023voicebox}, the system introduces two core components to enable environment-aware generation: an SES module (Fig.~\ref{fig: ses}) for disentangling speaker and environment information and environment conditioning mechanism in the TTS module (Fig.~\ref{fig: model_train}) for controllable recombination.

The SES module (Sec.~\ref{sec: ses}) decomposes an environmental speech signal into a clean speech component, noise component, and reverberation component, which are converted into conditions for the TTS module: mel-spectrograms $\mathbf{c}_{spk}$ and $\mathbf{c}_{noise}$ and an utterance-level reverberation embedding $\mathbf{c}_{reverb}$. 
The TTS module (Sec.~\ref{sec: tts}) generates the target environmental mel-spectrogram conditioned on these factors along with the text input, which is subsequently converted to a waveform by using a vocoder.

During training (Sec.~\ref{sec: pre-training}), the two modules are trained separately using simulated data with ground-truth signals, with a cross-speaker conditioning strategy to prevent speaker information leakage. 
When real-world data are available, the extended DAIEN-TTS can be further fine-tuned (Sec.~\ref{sec: finetuning}) by connecting the two modules into a full disentanglement-recombination pipeline.
 At inference (Sec.~\ref{sec: inference}), TCFG and SNR adaptation enable independent control over the speech, noise, and reverberation components.

\subsection{SES Module}
\label{sec: ses}
\begin{figure}[t!]
  \centering
  \includegraphics[width=\linewidth]{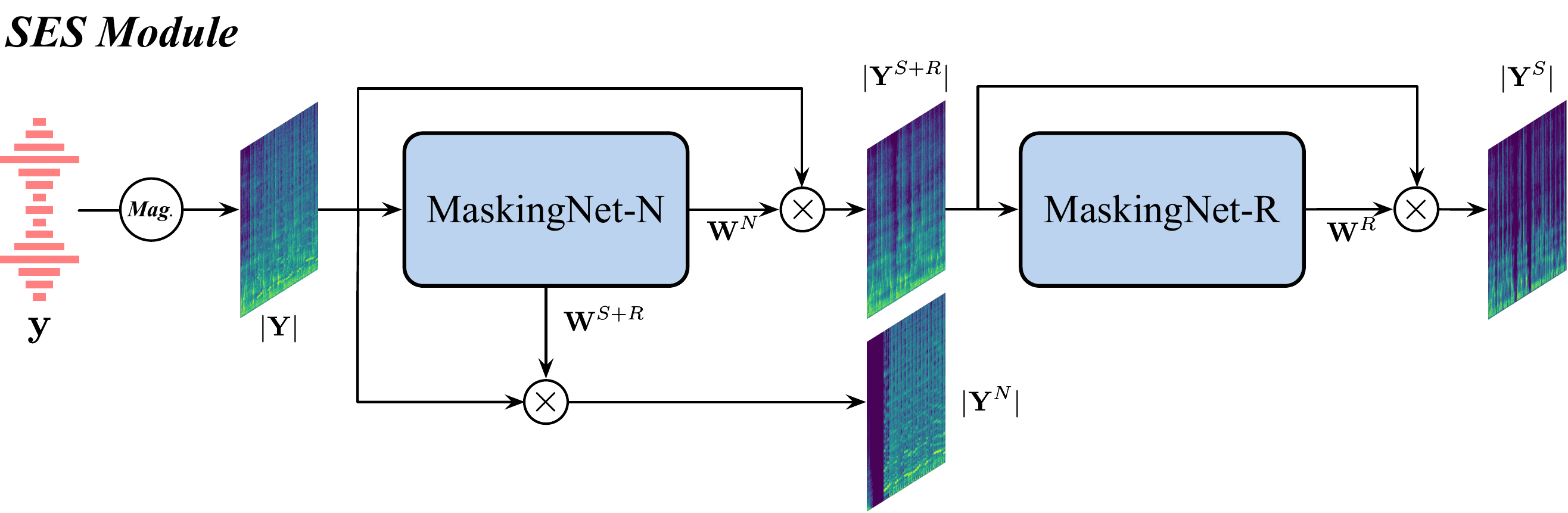}
  \caption{Overview of the SES module with two-stage noise and reverberation separation}
  \label{fig: ses}
\end{figure}

Time-frequency masking is a well-established technique for audio source separation and speech enhancement \cite{chandna2017monoaural, wang2018supervised}, where a mask is estimated and applied to the mixture spectrogram to extract or suppress target signal components. 
It has also been shown to be effective for suppressing convolutive distortions such as reverberation \cite{williamson2017time}.
 Since the downstream TTS module operates on mel-spectrograms derived from magnitude spectrograms and does not require phase information, our SES module is designed to execute masking in the magnitude spectrogram domain. 
As illustrated in Fig.~\ref{fig: ses}, the SES module adopts a cascaded two-stage design, where the first stage separates noise and the second stage separates reverberation.

In the first stage, the input environmental speech $\mathbf{y}$ is split into two components: a reverberant speech component and noise component. 
This is achieved by transforming $\mathbf{y}$ into a magnitude spectrogram $|\mathbf{Y}| \in \mathbb{R}^{F \times L}$ via the short-time Fourier transform, where $F$ and $L$ denote the number of frequency bins and time frames, respectively. 
A Transformer-based \cite{vaswani2017attention} masking network for noise separation (MaskingNet-N) then predicts two complementary separation masks $\mathbf{W}^{S+R}$ and $\mathbf{W}^N \in \mathbb{R}^{F \times L}$, which are applied to $|\mathbf{Y}|$ via element-wise multiplication:
\begin{equation}
|\mathbf{Y}^{S+R}| = |\mathbf{Y}| \odot \mathbf{W}^N, \quad |\mathbf{Y}^N| = |\mathbf{Y}| \odot \mathbf{W}^{S+R}.
\end{equation}

In the second stage, the reverberation is further separated from the reverberant speech. 
Another masking network (MaskingNet-R) predicts a reverberation mask $\mathbf{W}^R \in \mathbb{R}^{F \times L}$, and the clean speech magnitude spectrogram is obtained by
\begin{equation}
|\mathbf{Y}^S| = |\mathbf{Y}^{S+R}| \odot \mathbf{W}^R.
\end{equation}


The SES module thus outputs three disentangled components: $|\mathbf{Y}^S|$, $|\mathbf{Y}^N|$, and $\mathbf{W}^R$.
Note that unlike noise, which is additive and can be directly separated from the mixture as an independent signal, reverberation is a convolutive effect inseparable from the speech content in the spectro-temporal domain.
Therefore, rather than attempting to recover the RIR, we use the reverberation mask $\mathbf{W}^R$ as an implicit representation of the reverberation characteristics, as it encodes the spectro-temporal distortion introduced by the room acoustics.
How these three components are transformed into conditions for the TTS module is described in the next subsection.

\subsection{Environment-aware TTS Module}
\label{sec: tts}
Our TTS module is built upon F5-TTS \cite{chen2025f5}, which formulates zero-shot TTS as a text-guided speech infilling task \cite{le2023voicebox} under the conditional flow matching (CFM) framework \cite{lipman2023flow}.
Given a speech mel-spectrogram $\mathbf{x}_1$ and its corresponding text sequence $\mathbf{z}$, a random span mask $\mathbf{m}$ is applied to $\mathbf{x}_1$, where the unmasked portion serves as the speaker prompt.
Starting from a sampled Gaussian noise $\mathbf{x}_0 \sim \mathcal{N}(\mathbf{0}, \mathbf{I})$, the noisy input is constructed as $\psi_t(\mathbf{x}_0) = (1-t)\mathbf{x}_0 + t\mathbf{x}_1$, where $t$ is a randomly sampled flow step.
A Diffusion Transformer (DiT) \cite{peebles2023scalable} based network $\mathcal{V}_t$ is trained to predict the velocity field that transforms $\mathbf{x}_0$ towards $\mathbf{x}_1$, conditioned on the unmasked portion $(1-\mathbf{m}) \odot \mathbf{x}_1$ and the full text sequence $\mathbf{z}$, which includes both the transcription of the unmasked prompt and the text to be generated.
To extend this framework for environment-aware generation, we reformulate the task as a disentangled audio infilling problem, where the model simultaneously infills the masked speech and its corresponding acoustic environment, conditioned on the text and the disentangled speaker and environment representations.

The environment-aware TTS module takes three conditions as input: a frame-level speaker condition $\mathbf{c}_{spk}$, frame-level noise condition $\mathbf{c}_{noise}$, and utterance-level reverberation embedding $\mathbf{c}_{reverb}$. 
In the pretraining stage, the training target ($\mathbf{y}$ at the bottom of Fig.~\ref{fig: model_train} (a)) is a simulated environmental speech from a clean speech signal $\mathbf{s}$, noise signal $\mathbf{n}$, and RIR $\mathbf{h}$, as expressed with Eq.~(1).
 The $\mathbf{c}_{spk}$ and $\mathbf{c}_{noise}$ are obtained by directly converting $\mathbf{s}$ and $\mathbf{n}$ into mel-spectrograms. 
The $\mathbf{c}_{reverb}$ is extracted by convolving a different speaker's clean speech $\mathbf{s}'$ with the same RIR $\mathbf{h}$, computing the reverberation mask $\mathbf{W}^R$ from the resulting clean-reverberant pair, and passing it through an ECAPA-TDNN \cite{desplanques2020ecapa} encoder. 
This cross-speaker conditioning strategy ensures that $\mathbf{c}_{reverb}$ captures only the room acoustic characteristics without leaking the target speaker's identity.
 In the fine-tuning stage (Fig.~\ref{fig: model_train} (b)), the training target is real-world environmental speech, and the three conditions are provided by the SES module applied to the same input.

During training, random span masks of varying lengths $\mathbf{m}_{speech}$ and $\mathbf{m}_{noise}$ are independently applied to the disentangled clean speech mel-spectrogram $\mathbf{c}_{spk}$ and noise mel-spectrogram $\mathbf{c}_{noise}$, respectively.
The reverberation embedding $\mathbf{c}_{reverb}$ is an utterance-level representation, thus does not require masking.
During inference, the speaker and environment conditions are derived from two separate audio recordings that are generally of different lengths.
The clean speech condition is temporally aligned with the text sequence, the noise condition has no such constraint, and the reverberation embedding is utterance-level, thus independent of the prompt length.
Applying masks of independent lengths during training enables the model to accommodate this length mismatch at inference, avoiding truncation or padding of the environment condition that could lead to loss of environment information.

The three conditions are injected into the DiT blocks through different mechanisms, as depicted in the middle of Fig.~\ref{fig: model_train}. 
The unmasked speaker condition $(1-\mathbf{m}_{speech}) \odot c_{spk}$ is concatenated with the text embedding $\mathbf{z}$ and the noisy input $\psi_t(\mathbf{x}_0)$ along the feature dimension, following the original F5-TTS design. 
For the environment conditions, the reverberation embedding $c_{reverb}$ is first repeated along the frame dimension to match the length of the noise condition then added to the unmasked noise condition $(1-\mathbf{m}_{noise}) \odot c_{noise}$ to form a unified environment representation. 
This combined environment condition is introduced via a multi-head cross-attention layer inserted into each DiT block, enabling the model to jointly attend to frame-level noise patterns and global reverberation characteristics. 
With these conditions, the model learns to reconstruct the masked environmental speech mel-spectrogram $\mathbf{m}_{speech} \odot \mathbf{x}_1$, including both the personalized speech content and its corresponding acoustic environment.

\subsection{Pretraining on Simulated Data}
\label{sec: pre-training}

As illustrated in Fig.~\ref{fig: model_train} (a), we construct simulated environmental speech by mixing clean speech with noise and RIRs at various SNR levels, providing paired data with known ground-truth components for supervised training of both modules.

The SES module is pretrained independently with a combination of magnitude spectrogram and mel-spectrogram reconstruction losses over all three separated components:
\begin{equation}
\mathcal{L}_{\text{SES}} = \mathcal{L}_{\text{mag}} + \mathcal{L}_{\text{mel}},
\end{equation}
where
\begin{equation}
\begin{aligned}
\mathcal{L}_{\text{mag}} &= \left\| |\hat{\mathbf{Y}}^{S+R}| - |\mathbf{Y}^{S+R}| \right\|_2^2 \\
&+ \left\| |\hat{\mathbf{Y}}^N| - |\mathbf{Y}^N| \right\|_2^2 + \left\| |\hat{\mathbf{Y}}^S| - |\mathbf{Y}^S| \right\|_2^2,
\end{aligned}
\end{equation}
\begin{equation}
\begin{aligned}
&\mathcal{L}_{\text{mel}} = \left\| \text{Mel}(|\hat{\mathbf{Y}}^{S+R}|) - \text{Mel}(|\mathbf{Y}^{S+R}|) \right\|_1 \\
&+ \left\| \text{Mel}(|\hat{\mathbf{Y}}^N|) - \text{Mel}(|\mathbf{Y}^N|) \right\|_1 + \left\| \text{Mel}(|\hat{\mathbf{Y}}^S|) - \text{Mel}(|\mathbf{Y}^S|) \right\|_1,
\end{aligned}
\end{equation}
$|\hat{\mathbf{Y}}|$ and $|\mathbf{Y}|$ denote the predicted and corresponding ground-truth magnitude spectrogram, and $\text{Mel}(\cdot)$ denotes the log-mel-spectrogram conversion via a mel filter bank.

The TTS module is pretrained separately using the conditional flow matching objective:
\begin{align}
\mathcal{L}_{\mathrm{CFM}} = \Big \Vert &
\big[ \mathcal{V}_t\big(\psi_t(\mathbf{x}_0) \vert \mathbf{z},
(1 - \mathbf{m}_{speech}) \odot \mathbf{c}_{spk}, (1 - \mathbf{m}_{noise}) \notag \\
& \odot \mathbf{c}_{noise}, \mathbf{c}_{reverb}\big)
- (\mathbf{x}_1 - \mathbf{x}_0) \big]
\odot \mathbf{m}_{speech} \Big\Vert_2^2 \label{eq:cfm},
\end{align}
where $\mathcal{V}_t(\cdot)$ is the learned velocity field.
As described in Sec.~\ref{sec: tts}, the reverberation mask used for extracting $\mathbf{c}_{reverb}$ is computed from a different speaker's utterance to prevent speaker information leakage.
Specifically, for each training sample, we randomly select an utterance from a different speaker, truncate or repeat-pad it to match the target length, and convolve it with the same RIR applied to the original utterance.
This ensures that the reverberation mask retains only the room acoustic characteristics while discarding the target speaker's identity.
Noise and reverberation augmentations are independently applied to each clean speech sample with probabilities $p_{noise}$ and $p_{rir}$, respectively, resulting in four possible training conditions: clean speech, noisy speech, reverberant speech, and noisy-reverberant speech.
This ensures that the model learns to handle diverse environmental conditions while maintaining robust text-speech alignment from the clean speech cases.

\subsection{Fine-tuning on Real-world Data}
\label{sec: finetuning}

As illustrated in Fig.~\ref{fig: model_train} (b), to bridge the domain gap between simulated training data and real-world recordings, we fine-tune the extended DAIEN-TTS on real-world environmental speech data by connecting the SES and TTS modules into a full pipeline.
Unlike pretraining where the TTS module receives ground-truth conditions, during fine-tuning the three conditions $\mathbf{c}_{spk}$, $\mathbf{c}_{noise}$, and $\mathbf{c}_{reverb}$ are directly provided by the SES module applied to the real-world environmental speech.
The TTS module is then trained to reconstruct the same input environmental speech using $\mathcal{L}_{\mathrm{CFM}}$.

Several components are frozen during fine-tuning.
The SES module is frozen as real-world environmental speech lacks paired clean speech references, making supervised optimization of $\mathcal{L}_{\text{SES}}$ infeasible.
The text encoder is frozen to preserve the text-speech alignment acquired during pretraining.
The ECAPA-TDNN reverb encoder is frozen because ground-truth RIRs are unavailable for real-world recordings, precluding the application of the cross-speaker conditioning strategy; updating the reverb encoder without this constraint would risk speaker information leakage through the environment branch.
The remaining parameters, including the DiT blocks and environment cross-attention layers, are updated to adapt to the acoustic characteristics of real-world environments.

\subsection{Inference}
\label{sec: inference}

\begin{figure}[t!]
  \centering
  \includegraphics[width=0.75\linewidth]{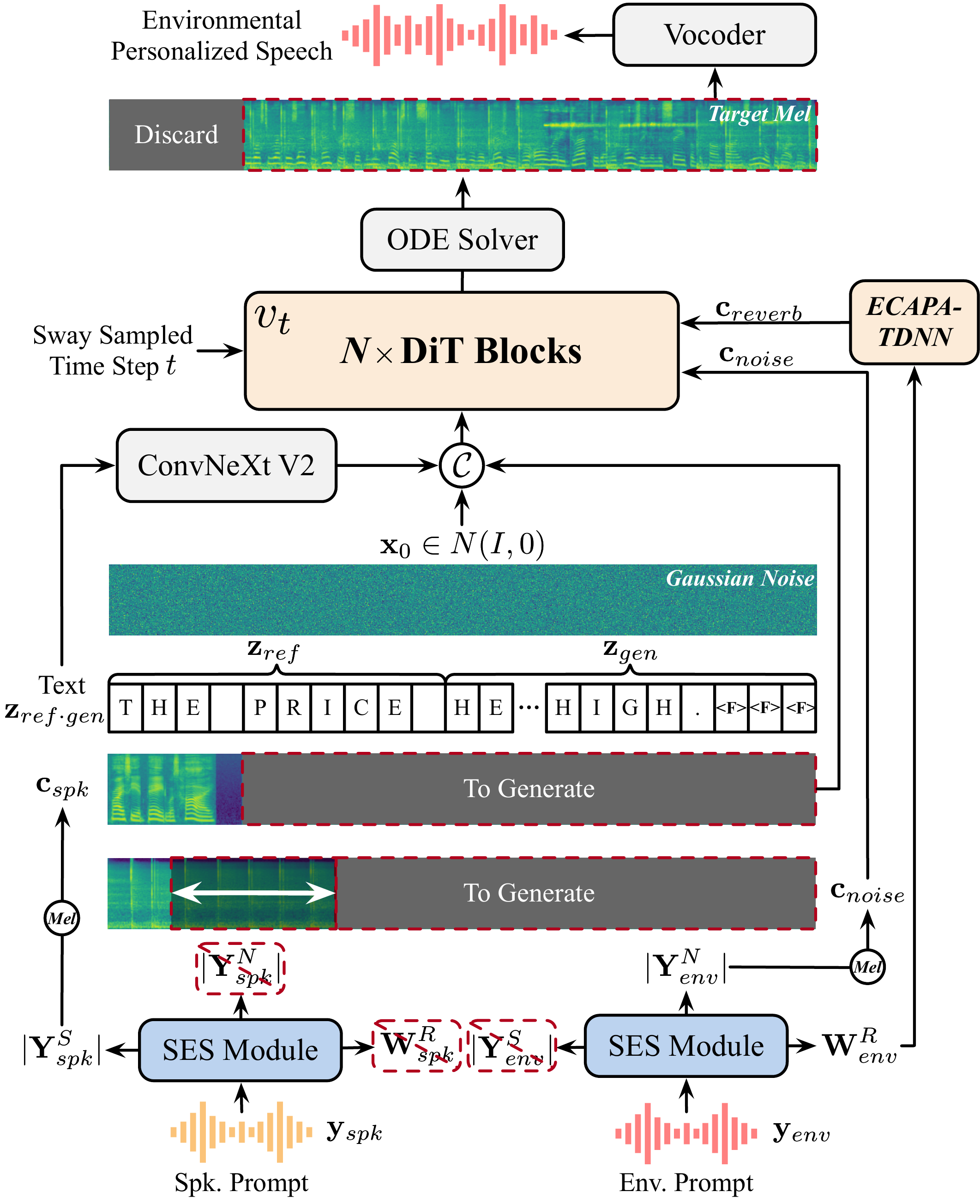}
  \caption{Inference process of the extended DAIEN-TTS. Given text 
  $\mathbf{z}$, a speaker prompt $\mathbf{y}_{spk}$, and environment prompt $\mathbf{y}_{env}$, the SES module extracts the speaker and environment conditions from the speaker prompt and environment prompt, respectively, for generation.}
  \label{fig: model_infer}
\end{figure}

The inference process is illustrated in Fig.~\ref{fig: model_infer}. 
Given a speaker audio prompt $\mathbf{y}_{spk}$ and environment audio prompt $\mathbf{y}_{env}$, the SES module is applied to each prompt separately: from $\mathbf{y}_{spk}$, we extract the clean speech component $|\mathbf{Y}^S_{spk}|$ and convert it to the speaker condition $\mathbf{c}_{spk}$; from $\mathbf{y}_{env}$, we extract the noise component $|\mathbf{Y}^N_{env}|$ and reverberation mask $\mathbf{W}^R_{env}$, which are converted to $\mathbf{c}_{noise}$ and $\mathbf{c}_{reverb}$, respectively. 
With the transcription of the speaker prompt $\mathbf{z}_{ref}$ and the text to be generated $\mathbf{z}_{gen}$, we concatenate them as $\mathbf{z}_{ref \cdot gen}$ and use an ordinary differential equation (ODE) solver to generate the target mel-spectrogram. 
Starting from sampled Gaussian noise $\mathbf{x}_0 \sim \mathcal{N}(\mathbf{0}, \mathbf{I})$, the ODE solver integrates towards the target $\mathbf{x}_1$ according to
\begin{equation}
d\psi_t(\mathbf{x}_0) / dt = \mathcal{V}_t\big(\psi_t(\mathbf{x}_0), \mathbf{z}_{ref \cdot gen}, \mathbf{c}_{spk}, \mathbf{c}_{noise}, \mathbf{c}_{reverb}\big).
\end{equation}
After generation, the prompt region corresponding to $\mathbf{z}_{ref}$ is discarded, and the remaining generated mel-spectrogram is converted to a waveform by using the vocoder.

The above inference procedure generates environmental speech with the specified speaker and environment conditions but offers limited control over the contribution of each environmental component to the synthesized output.
To enhance the controllability at inference, we further introduce two complementary strategies: TCFG for independently adjusting the strength of the speech, noise, and reverberation components, and SNR adaptation for aligning the noise level of the synthesized speech with that of the environment prompt.

\subsubsection{Triple Classifier-free Guidance}
F5-TTS uses CFG \cite{ho2022classifier} to balance generation fidelity and diversity by treating the text and speaker conditions as a unified guidance term. 
In our prior work \cite{lu2026daien}, we adopted a DCFG \cite{lee2024voiceldm} mechanism to separately guide the speech and environment components. 
In this work, since the environment condition is further decomposed into noise and reverberation, we extended DCFG to TCFG for independent control over the speech, noise, and reverberation components.
 Let $\mathbf{c}_{speech} = \{\mathbf{z}, \mathbf{c}_{spk}\}$ denote the unified speech condition combining the text sequence and speaker condition. 
The TCFG is formulated as
\begin{equation}
\begin{aligned}
&\mathcal{V}_{t,\text{TCFG}} = \mathcal{V}_t(\psi_t(\mathbf{x}_0), \mathbf{c}_{speech}, \mathbf{c}_{noise}, \mathbf{c}_{reverb}) \\
&+ \alpha_{speech} \left[ \mathcal{V}_t(\psi_t(\mathbf{x}_0), \mathbf{c}_{speech}, \varnothing, \varnothing) - \mathcal{V}_t(\psi_t(\mathbf{x}_0), \varnothing, \varnothing, \varnothing) \right] \\
&+ \alpha_{noise} \left[ \mathcal{V}_t(\psi_t(\mathbf{x}_0), \varnothing, \mathbf{c}_{noise}, \varnothing) - \mathcal{V}_t(\psi_t(\mathbf{x}_0), \varnothing, \varnothing, \varnothing) \right] \\
&+ \alpha_{reverb} \left[ \mathcal{V}_t(\psi_t(\mathbf{x}_0), \varnothing, \varnothing, \mathbf{c}_{reverb}) - \mathcal{V}_t(\psi_t(\mathbf{x}_0), \varnothing, \varnothing, \varnothing) \right],
\end{aligned}
\end{equation}
where $\varnothing$ denotes the null condition, and $\alpha_{speech}$, $\alpha_{noise}$, and $\alpha_{reverb}$ are the guidance scales for the speech, noise, and reverberation components, respectively.
The derivation follows naturally from the DCFG formulation \cite{lee2024voiceldm} by extending to three independent conditions.
Adjusting these three scales independently enables fine-grained control over the contribution of each component to the synthesized environmental speech.

\subsubsection{SNR Adaptation}
The SNR of the environment prompt characterizes the relative prominence of the background noise in the acoustic scene, thus an important attribute to be preserved in the synthesized speech. 
However, since the speaker and environment conditions are extracted from separate audio prompts with potentially different energy levels, a direct combination of these conditions may result in an SNR discrepancy between the synthesized speech and environment prompt. 
To mitigate this, we introduce an SNR adaptation strategy that rescales the noise condition such that the SNR of the synthesized speech is consistent with that of the environment prompt.

Considering the environmental speech $\mathbf{y}$ is composed of its reverberant speech component $\mathbf{y}^{S+R}$ and noise component $\mathbf{y}^N$, such that $\mathbf{y} = \mathbf{y}^{S+R} + \mathbf{y}^N$, the SNR can be expressed as
\begin{equation}
\text{SNR} = 10 \log_{10} \frac{\| \mathbf{y}^{S+R} \|_2^2}{\| \mathbf{y}^N \|_2^2} = 10 \log_{10} \frac{\| \mathbf{Y}^{S+R} \|_2^2}{\| \mathbf{Y}^N \|_2^2},
\end{equation}
where the second equality follows from Parseval's theorem.
To match the SNR of the synthesized speech with that of the environment prompt, we require
\begin{equation}
10 \log_{10} \frac{\| \mathbf{Y}^{S+R}_{spk} \|_2^2}{\| \mathbf{Y}^N_{env} \cdot \text{Scale} \|_2^2} = 10 \log_{10} \frac{\| \mathbf{Y}^{S+R}_{env} \|_2^2}{\| \mathbf{Y}^N_{env} \|_2^2},
\end{equation}
where $\mathbf{Y}^{S+R}_{spk}$ denotes the separated reverberant speech magnitude spectrogram of the speaker prompt, and $\mathbf{Y}^{S+R}_{env}$ and $\mathbf{Y}^N_{env}$ denote the separated reverberant speech and noise magnitude spectrograms of the environment prompt, respectively.
Solving for the scaling factor yields
\begin{equation}
\text{Scale} = \sqrt{\frac{\| \mathbf{Y}^{S+R}_{spk} \|_2^2}{\| \mathbf{Y}^{S+R}_{env} \|_2^2}}.
\end{equation}
During inference, $|\mathbf{Y}^N_{env}|$ is multiplied by this factor prior to mel-spectrogram conversion, ensuring that the resulting noise condition reflects the target SNR of the environment prompt relative to the speech energy of the speaker prompt.

\section{Experimental Setup}

\subsection{Datasets}

Our experiments involve two types of data: simulated environmental speech for pretraining and evaluation under controlled conditions, and real-world environmental speech for fine-tuning and evaluation under realistic conditions.

\subsubsection{Simulated Data}
For pretraining, we used the LibriTTS corpus \cite{zen2019libritts}, which contains 580 hours of high-quality English speech data.
To construct simulated environmental speech, the clean speech was mixed with noise samples from the DNS-Challenge noise set \cite{dubey2022icassp}, which comprises approximately 180 hours of environmental audio sourced from AudioSet \cite{gemmeke2017audio} and FreeSound \cite{fonseca2017freesound}.
RIRs were sampled from the DNS-Challenge RIR set \footnote{\url{https://www.openslr.org/26/}, \url{https://www.openslr.org/28/}}, containing 20,248 samples.
The SNR for noise mixing was uniformly sampled between $-5$ and $15$ dB, and the noise and RIR augmentations were independently applied with probabilities $p_{noise}$ and $p_{rir}$, both set to 0.5.

For evaluation under simulated conditions, we constructed a test set based on the Seed-TTS test-en set\footnote{\url{https://github.com/BytedanceSpeech/seed-tts-eval}}, which originally provides the text to be synthesized, a speaker reference audio with its transcription, and a corresponding human-recorded ground truth for each sample.
To evaluate environment-aware TTS, we extended each sample with an environment reference audio by randomly selecting an utterance from a different speaker.
Both the speaker and environment reference audios were augmented with independently sampled noise and RIRs, with SNRs uniformly distributed between $0$ and $20$ dB.
The noise clips were sourced from SoundBible\footnote{\url{https://soundbible.com}} and the RIRs from the SRIRACHA dataset \cite{pelling2025sriracha}.
To construct the ground truth for the environmental personalized speech, the original clean ground-truth recording was first convolved with the same RIR applied to the environment reference audio.
For the noise component, since the TTS model is expected to continue the background noise from the environment prompt, the noise signal appended to the ground truth was taken from the segment immediately following the portion used in the environment reference audio, preserving temporal continuity.
All data were constructed at a sampling rate of 24 kHz.

\subsubsection{Real-world Data}
For fine-tuning, we used the TITW-Hard subset of the TITW (TTS in the Wild) dataset \cite{jung2025text}, which is derived from VoxCeleb 1 \cite{nagrani2017voxceleb} through an automated transcription, segmentation, and selection pipeline, containing 273,493 utterances totaling 184.57 hours of English speech recorded under diverse real-world acoustic conditions.
Since the original TITW data are recorded at 16 kHz, all audio was upsampled to 24 kHz to match the sampling rate of the pretraining data.

For evaluation under real-world conditions, we used text from the LibriTTS \textit{test-clean} set, while the speaker and environment reference audios were selected from the TITW-Hard test set.
To ensure distinct acoustic conditions between the two references, we scored all TITW test samples using DNSMOS \cite{reddy2022dnsmos} and selected speaker prompts from samples with background-noise quality (BAK) scores above 3.5, indicating relatively clean recordings, and environment prompts from samples with BAK scores below 2.5, indicating prominent background noise or reverberation.

\subsection{Model Configuration}

In the SES module, both MaskingNet-N and MaskingNet-R consist of an input convolutional layer, 8 Transformer blocks with 16 attention heads, an embedding dimension of 1024, and feed-forward network (FFN) dimension of 2048, followed by output convolutional layers. 
For the TTS module, we adopted the same DiT architecture as F5-TTS, with each DiT block inserted by a cross-attention layer configured with 16 attention heads for environment conditioning. 
The ECAPA-TDNN \cite{desplanques2020ecapa} encoder for reverberation embedding extraction is jointly trained with the TTS module during pretraining, and frozen during fine-tuning to prevent speaker information leakage.
 All audio was represented as 100-dimensional log mel-filterbank features at a sampling rate of 24 kHz with a hop length of 256.

Both SES and TTS modules were pretrained for 600k steps on 4 NVIDIA A100 80G GPUs, using the AdamW optimizer with a peak learning rate of $7.5 \times 10^{-5}$, linearly warmed up for 20k steps and linearly decayed thereafter, and a batch size of 120,000 audio frames.
 During training, a random 70 to 100\% of mel frames are independently masked for both the speech and noise conditions, while the reverberation condition is always provided in full. 
For CFG training, the speech, noise, and reverberation conditions are each independently dropped with a probability of 0.3, and the speech condition together with the text input is jointly dropped with an additional probability of 0.2.
 For fine-tuning on real-world data, the learning rate was reduced to $1 \times 10^{-5}$ with a warm-up of 2k steps, and the system was fine-tuned for 50k steps. 
During inference, the TCFG guidance scales were set to $\alpha_{speech} = 1$, $\alpha_{noise} = 3$, and $\alpha_{reverb} = 6$ unless otherwise specified, and the number of function evaluations for the ODE solver was set to 32. 
For the vocoder, we trained a Vocos \cite{siuzdak2024vocos} model on environmental speech data constructed with the same noise and RIR augmentation pipeline as the pretraining data, enabling it to reconstruct environmental speech waveforms from mel-spectrograms. 
\footnote{Audio samples: \href{https://yxlu-0102.github.io/DAIEN-TTS/journal}{https://yxlu-0102.github.io/DAIEN-TTS/journal}.}

\subsection{Baseline and Compared Systems}

We evaluated the following systems in our experiments:
\begin{itemize}
\item \textbf{F5-TTS} \cite{chen2025f5}: The base F5-TTS model trained on the LibriTTS dataset for 600k steps, representing a standard zero-shot TTS system without environment awareness.
\item \textbf{DAIEN-TTS-N} \cite{lu2026daien}: The conference version of DAIEN-TTS, which only models background noise without reverberation and trained exclusively on simulated data.
\item \textbf{DAIEN-TTS-NR}: The extended DAIEN-TTS that jointly models background noise and reverberation, pretrained on simulated data.
\item \textbf{DAIEN-TTS-FT}: The extended DAIEN-TTS fine-tuned on real-world speech data from the TITW-Hard dataset.
\end{itemize}
For a fair comparison, all systems share the same LibriTTS pretraining data and Vocos vocoder.

\subsection{Evaluation Metrics}

Following standard zero-shot TTS evaluation practice, we assessed the synthesized speech in terms of naturalness, intelligibility, and speaker similarity. 
Given the environment-aware nature of our task, we further evaluated environmental fidelity to measure how well the synthesized speech reproduces the target acoustic environment. 
Both objective metrics and subjective listening tests were used for evaluation.

\subsubsection{Objective Metrics}
Speech intelligibility was assessed by computing the word error rate (WER) using Whisper-large-v3 \cite{radford2023robust} to transcribe the synthesized speech and comparing the transcription against the target text.
A lower WER indicates higher intelligibility and more accurate text-speech alignment.

Speaker similarity was quantified by computing the cosine similarity between speaker embeddings extracted from the synthesized speech and the speaker prompt using the WavLM-large-based speaker verification model \cite{chen2022large}, denoted as SECS (Speaker Embedding Cosine Similarity).
Since speaker verification models are typically trained on data with diverse acoustic conditions, the extracted embeddings are reasonably robust to environmental variations, enabling speaker similarity to be evaluated even when the synthesized speech contains background noise or reverberation.

Since there are no established objective metrics for evaluating environmental fidelity in environment-aware TTS, we propose to adopt two complementary metrics for this purpose.
Contrastive language-audio pretraining (CLAP) similarity \cite{elizalde2023clap} measures the pairwise cosine similarity between CLAP embeddings of the synthesized speech and the environment prompt. 
CLAP is trained via contrastive learning on large-scale audio-text pairs, and its audio encoder captures high-level acoustic scene attributes such as background sound events and overall acoustic characteristics, rather than relying on waveform-level matching.
A higher CLAP similarity indicates better environmental consistency at the sample level.
We further use the Fréchet audio distance (FAD) \cite{kilgour2019frechet} computed using VGGish \cite{hershey2017cnn} embeddings to evaluate environmental fidelity at the distribution level.
FAD computes the Fréchet distance between Gaussian distributions fitted to the embeddings of the synthesized and reference audio sets, providing a holistic measure of how well the overall distribution of acoustic environments is reproduced.
A lower FAD indicates better environmental fidelity.

\subsubsection{Subjective Metrics}
We conducted three types of listening tests.
A mean opinion score (MOS) test was conducted to assess the overall naturalness of the synthesized speech, including both the speech content and acoustic environment, on a 1--5 scale with 0.5-point resolution.
A speaker similarity MOS (SSMOS) test was conducted by providing the speaker prompt as a reference and asking listeners to rate how closely the synthesized speech matches the target speaker's voice, focusing on speaker identity while ignoring the acoustic environment.
An environment similarity MOS (ESMOS) test was conducted by providing the environment prompt as a reference and asking listeners to rate the consistency in terms of background noise, non-speech sound events, and reverberation, while ignoring speaker identity and speech content.
All listening tests were crowd-sourced via Amazon Mechanical Turk with more than 30 native English-speaking raters, each rating 20 samples per test.
We report the mean scores with 95\% confidence intervals (CI).

\section{Results and Analysis}

\subsection{Environment-robust Generation}

We first evaluated the ability of each system to synthesize clean personalized speech when given a silence environment prompt, under two speaker prompt conditions: clean and simulated noisy-reverberant. 
In this scenario, the system is expected to disentangle the environmental factors from the speaker prompt and produce clean, natural speech. 
For the MOS test in this scenario, listeners were asked to evaluate both the naturalness and acoustic cleanliness of the synthesized speech, with any residual noise or reverberation artifacts considered as quality degradation. 
For the SSMOS test, the clean version of the speaker prompt was provided as the reference since paired clean-environmental speech is available in the simulated test set. 
Table~\ref{tab:robust} presents the evaluation results, where the ground-truth recordings and their vocoder-resynthesized versions serve as upper bounds.

Under the clean speaker prompt condition, all TTS systems achieved comparable performance across all metrics. 
The DAIEN-TTS variants outperformed F5-TTS in WER and SECS, which can be attributed to the data augmentation effect introduced by training on paired clean-environmental speech data. 
The MOS and SSMOS scores of all systems were close, with overlapping confidence intervals, indicating that the additional environment modeling in DAIEN-TTS does not degrade synthesis quality when clean prompts are provided.

Under the simulated noisy-reverberant speaker prompt condition, clear performance differences emerged among the systems. 
F5-TTS, which is trained exclusively on clean speech, exhibited substantial degradation in both intelligibility and speaker similarity compared with the clean prompt condition, confirming the sensitivity of ICL-based TTS to adverse acoustic conditions in the speaker prompt. 
DAIEN-TTS-N showed a similar trend, as its noise-only separation module cannot adequately remove the reverberation component from the speaker prompt. 
In contrast, DAIEN-TTS-NR achieved the best performance across all metrics, with its MOS score comparable to the ground-truth reference, demonstrating that joint modeling of noise and reverberation is essential for robust synthesis under adverse acoustic conditions.

\begin{table}[t]
\caption{Objective and subjective evaluation results of environment-robust generation on the Seed-TTS test-en set}
\label{tab:robust}
\centering
\resizebox{\columnwidth}{!}{%
\begin{tabular}{lcccc}
\toprule
System & WER (\%) $\downarrow$ & SECS $\uparrow$ & MOS (CI) $\uparrow$ & SSMOS (CI) $\uparrow$ \\
\midrule
\multicolumn{5}{c}{\textit{Clean Speaker Prompt}} \\
\midrule
Ground Truth & 2.14 & 0.734 & 3.91 ($\pm$ 0.08) & 3.52 ($\pm$ 0.09) \\
Vocoder & 2.18 & 0.698 & -- & -- \\ [3pt]
\hdashline \\[-6pt]
F5-TTS & 2.30 & 0.590 & 3.86 ($\pm$ 0.08) & 3.46 ($\pm$ 0.09) \\
DAIEN-TTS-N & 2.07 & 0.617 & \textbf{3.91} ($\pm$ 0.08) & \textbf{3.49} ($\pm$ 0.08) \\
DAIEN-TTS-NR & \textbf{2.03} & \textbf{0.628} & 3.90 ($\pm$ 0.08) & 3.48 ($\pm$ 0.08) \\
\midrule
\multicolumn{5}{c}{\textit{Simulated Noisy-reverberant Speaker Prompt}} \\
\midrule
Ground Truth & 2.14 & 0.734 & 3.95 ($\pm$ 0.07) & 3.43 ($\pm$ 0.08) \\
Vocoder & 2.18 & 0.698 & -- & -- \\
F5-TTS & 2.80 & 0.317 & 3.75 ($\pm$ 0.08) & 3.19 ($\pm$ 0.09) \\
DAIEN-TTS-N & 2.75 & 0.380 & 3.77 ($\pm$ 0.07) & 3.30 ($\pm$ 0.08) \\
DAIEN-TTS-NR & \textbf{2.24} & \textbf{0.486} & \textbf{3.91} ($\pm$ 0.08) & \textbf{3.37} ($\pm$ 0.08) \\
\bottomrule
\end{tabular}
}
\end{table}

\begin{table*}[t]
\caption{Objective and subjective evaluation results of environment-aware generation on the Seed-TTS test-en set (simulated prompts) and TITW test set (real-world prompts)}
\label{tab:aware}
\centering
\begin{tabular}{lccccccc}
\toprule
System & WER (\%) & SECS $\uparrow$ & CLAP SIM $\uparrow$ & FAD (VGGish) $\downarrow$ & MOS (CI) $\uparrow$ & SSMOS (CI) $\uparrow$ & ESMOS (CI) $\uparrow$ \\
\midrule
\multicolumn{8}{c}{\textit{Simulated Noisy-reverberant Prompts}} \\
\midrule
Ground Truth & 15.41 & 0.530 & 0.851 & 0.10 & 3.40 ($\pm$ 0.08) & 3.58 ($\pm$ 0.09) & 3.59 ($\pm$ 0.07) \\
Vocoder & 15.80 & 0.493 & 0.848 & 0.13 & -- & -- & -- \\ [3pt]
\hdashline \\[-6pt]
F5-TTS & 2.80 & 0.317 & 0.685 & 8.56 & 3.19 ($\pm$ 0.08) & 3.41 ($\pm$ 0.09) & 3.27 ($\pm$ 0.09) \\
DAIEN-TTS-N & 10.85 & 0.325 & 0.743 & 5.73 & 3.27 ($\pm$ 0.08) & 3.48 ($\pm$ 0.09) & 3.45 ($\pm$ 0.08) \\
DAIEN-TTS-NR & 13.05 & \textbf{0.361} & \textbf{0.770} & \textbf{2.11} & \textbf{3.38} ($\pm$ 0.08) & \textbf{3.53} ($\pm$ 0.08) & \textbf{3.55} ($\pm$ 0.08) \\
\midrule
\multicolumn{8}{c}{\textit{Real-world Environmental Prompts}} \\
\midrule
F5-TTS & 12.42 & 0.481 & 0.588 & 8.37 & 3.40 ($\pm$ 0.08) & 3.82 ($\pm$ 0.08) & 3.46 ($\pm$ 0.09) \\
DAIEN-TTS-N & 16.07 & 0.447 & 0.618 & 4.92 & 3.29 ($\pm$ 0.08) & 3.71 ($\pm$ 0.09) & 3.55 ($\pm$ 0.08) \\
DAIEN-TTS-NR & 10.05 & 0.457 & 0.612 & 5.33 & 3.34 ($\pm$ 0.08) & 3.74 ($\pm$ 0.08) & 3.58 ($\pm$ 0.08) \\
DAIEN-TTS-FT & 10.65 & \textbf{0.486} & \textbf{0.634} & \textbf{3.62} & 3.38 ($\pm$ 0.08) & \textbf{3.84} ($\pm$ 0.09) & \textbf{3.62} ($\pm$ 0.08) \\
\bottomrule
\end{tabular}
\end{table*}

\subsection{Environment-aware Generation}

We next evaluated the ability of each system to synthesize environmental personalized speech that reproduces the acoustic environment specified by the environment prompt. 
Experiments were conducted under two conditions: simulated speaker and environment prompts and real-world prompts. 
In contrast to the environment-robust scenario, listeners in the MOS test were instructed to evaluate the overall naturalness of the environmental speech, including both the speech and acoustic environment, without penalizing the presence of background noise and reverberation. 
For the SSMOS test under the simulated condition, the clean version of the speaker prompt was provided as the reference since paired clean-environmental speech is available. 
Under the real-world condition, the original speaker prompt was used as the reference, as paired clean recordings are unavailable.
 Table~\ref{tab:aware} presents the evaluation results.

\subsubsection{Simulated Prompts}
Under the simulated prompt condition, the ground-truth recordings and their vocoder-resynthesized versions serve as upper bounds. 
Note that WER in this scenario should be interpreted with caution. 
The ground-truth environmental speech already exhibits a WER of 15.41\%, significantly higher than the 2.14\% observed for clean speech in Table~\ref{tab:robust}, indicating that the ASR system is substantially affected by the presence of noise and reverberation. 
Similarly, higher WERs for the environment-aware systems may reflect the expected presence of background noise and reverberation in the synthesized speech, rather than a degradation in synthesis quality. 
F5-TTS achieved the lowest WER as it lacks an environment conditioning mechanism and does not actively reproduce the target acoustic environment.

Among the environment-aware systems, DAIEN-TTS-N partially reproduced the background noise but lacked reverberation modeling, resulting in moderate environmental fidelity (CLAP: 0.743, FAD: 5.73). 
DAIEN-TTS-NR achieved the best environmental fidelity (CLAP: 0.770, FAD: 2.11), substantially narrowing the gap to the vocoder upper bound (FAD: 0.13), while also attaining the highest speaker similarity (SECS: 0.361) among all TTS systems.

The subjective evaluation results further corroborate these findings. 
DAIEN-TTS-NR achieved the highest scores across all three perceptual dimensions (MOS: 3.38, SSMOS: 3.53, ESMOS: 3.55), with its ESMOS approaching the ground-truth reference (3.59), indicating that the reconstructed acoustic environment is perceptually close to the target. 
F5-TTS received the lowest MOS (3.19) and ESMOS (3.27), as it is trained on clean speech and struggles to faithfully reconstruct the heavy noise and reverberation present in the speaker prompt, resulting in degraded environmental quality despite partially preserving some background characteristics. 
These results confirm that joint modeling of noise and reverberation significantly improves the fidelity of environment reconstruction in both objective and subjective evaluations.

\subsubsection{Real-world Prompts}
Under the real-world prompt condition, we further included the fine-tuned variant DAIEN-TTS-FT to evaluate the effectiveness of real-world data adaptation. 
Note that the WERs under the real-world condition were generally higher than those under the simulated condition for all systems, partly because the transcriptions of the real-world speaker prompts were obtained via ASR and may contain errors, which propagate through the text input and affect the synthesized speech.

As expected, F5-TTS yielded the lowest CLAP similarity (0.588) and highest FAD (8.37), as it lacks the ability to reproduce acoustic environments by design.
 However, it achieved a relatively high MOS (3.40) and the highest SSMOS (3.82) under this condition, as the speaker prompts were selected to be relatively clean (DNSMOS BAK $>$ 3.5), resulting in higher speech quality despite poor environment reconstruction.

DAIEN-TTS-N achieved higher environmental fidelity (CLAP: 0.618, FAD: 4.92), but its noise-only separation module struggled with the complex acoustic conditions in real-world recordings, leading to noticeable degradation in speech intelligibility. 
The additional reverberation modeling in DAIEN-TTS-NR did not yield a clear advantage over DAIEN-TTS-N in environmental fidelity under this condition. 
An analysis of the TITW test samples revealed that the acoustic environments are predominantly characterized by background noise, with reverberation playing a less prominent role, which limits the benefit of joint noise-reverberation modeling.

After fine-tuning on real-world data, DAIEN-TTS-FT achieved improvements across all metrics, attaining the best speaker similarity (SECS: 0.486), environmental fidelity (CLAP: 0.634, FAD: 3.62), and subjective scores (MOS: 3.38, SSMOS: 3.84, ESMOS: 3.62), demonstrating the effectiveness of the two-stage training strategy in bridging the simulated-to-real domain gap.

\subsection{Analysis Experiment}

\subsubsection{Effect of Environment Conditions}
To further investigate the behavior of each system under different types of acoustic environments, we evaluated environment-aware generation with noise-only and reverb-only environment prompts on the simulated test set.
Table~\ref{tab:env_conditions} presents the results, complementing the noisy-reverberant results in Table~\ref{tab:aware}.

Under the noise-only condition, F5-TTS yielded the lowest CLAP similarity and highest FAD, consistent with its inability to reproduce acoustic environments. 
Both DAIEN-TTS-N and DAIEN-TTS-NR achieved substantially better environmental fidelity, with DAIEN-TTS-N attaining slightly higher CLAP similarity as it is specifically designed for noise-only modeling. 
The marginal difference between the two systems indicates that the additional reverberation modeling in DAIEN-TTS-NR does not significantly interfere with noise disentanglement and reconstruction in noise-dominated scenarios.

Under the reverb-only condition, a substantial performance gap emerged. 
DAIEN-TTS-N, which lacks reverberation modeling, exhibited poor environmental fidelity (FAD: 5.06), as its noise-only separation module fails to properly handle reverberant speech. 
In contrast, DAIEN-TTS-NR achieved notably better environmental fidelity (FAD: 2.86) and speaker similarity (SECS: 0.400 vs. 0.317), confirming the necessity of explicit reverberation modeling.

We further analyzed the characteristics of the two environmental fidelity metrics by computing both between the noisy-reverberant environment prompt and ground-truth recordings under four conditions, as shown in Table~\ref{tab:metric_analysis}.
The results reveal that the two metrics capture different aspects of environmental fidelity: CLAP similarity, based on high-level audio scene embeddings, reflects the overall acoustic similarity, while FAD, computed at the distribution level, is particularly effective at capturing reverberation characteristics, as evidenced by its sharp decrease from 17.05 (clean) to 2.78 (reverb-only).

The results from Table~\ref{tab:metric_analysis} also help explain the observation in Table~\ref{tab:env_conditions}, which shows that F5-TTS achieved a relatively high CLAP similarity (0.814) under the reverb-only condition.
As an ICL-based system, F5-TTS tends to preserve the reverberation from the speaker prompt, contributing to a moderate CLAP similarity, but its high FAD (7.11) reveals that the reverberation characteristics do not faithfully match those of the environment prompt.
The complementary nature of the two metrics justifies our adoption of both for evaluating environmental fidelity.

These results, together with the noisy-reverberant results in Table~\ref{tab:aware}, indicate that DAIEN-TTS-NR maintains robust performance across diverse environment conditions, while DAIEN-TTS-N is limited to noise-dominated scenarios.

\begin{table}[t]
\caption{Environment-aware generation results under noise-only and reverb-only conditions on the simulated test set}
\label{tab:env_conditions}
\centering
\resizebox{\columnwidth}{!}{%
\begin{tabular}{lcccc}
\toprule
System & WER (\%)  & SECS $\uparrow$ & CLAP SIM $\uparrow$ & FAD (VGGish) $\downarrow$ \\
\midrule
\multicolumn{5}{c}{\textit{Noise-only Prompts}} \\
\midrule
Ground Truth & 3.38 & 0.699 & 0.788 & 0.13 \\
Vocoder & 3.48 & 0.653 & 0.784 & 0.18 \\
\hdashline \\[-6pt]
F5-TTS & 2.58 & 0.506 & 0.612 & 3.83 \\
DAIEN-TTS-N & 3.03 & \textbf{0.543} & \textbf{0.735} & 1.54 \\
DAIEN-TTS-NR & 2.90 & 0.528 & 0.716 & \textbf{1.25} \\
\midrule
\multicolumn{5}{c}{\textit{Reverb-only Prompts}} \\
\midrule
Ground Truth & 5.29 & 0.605 & 0.873 & 0.04 \\
Vocoder & 5.20 & 0.574 & 0.874 & 0.08 \\
\hdashline \\[-6pt]
F5-TTS & 2.68 & 0.391 & 0.814 & 7.11 \\
DAIEN-TTS-N & 17.56 & 0.317 & 0.770 & 5.06 \\
DAIEN-TTS-NR & 13.53 & \textbf{0.400} & 0.794 & \textbf{2.86} \\
\bottomrule
\end{tabular}%
}
\end{table}
\begin{table}[t]
\caption{CLAP similarity and FAD between the noisy-reverberant environment prompt and ground-truth recordings under different acoustic conditions}
\label{tab:metric_analysis}
\centering
\begin{tabular}{lcc}
\toprule
Evaluation Set & CLAP SIM $\uparrow$ & FAD (VGGish) $\downarrow$ \\
\midrule
Clean & 0.540 & 17.05 \\
Noise-Only & 0.672 & 12.72 \\
Reverb-Only & 0.729 & 2.78 \\
Noisy-Reverberant & 0.851 & 0.10 \\
\bottomrule
\end{tabular}
\end{table}

\subsubsection{Effect of TCFG Scales}

To demonstrate the independent controllability of the TCFG mechanism, we varied $\alpha_{noise}$ and $\alpha_{reverb}$ separately while keeping the other scales fixed and evaluated the results on the simulated noisy reverberant test set. 
Table~\ref{tab:tcfg} presents the results. 
The guidance scales used in the main experiments ($\alpha_{speech}=1$, $\alpha_{noise}=3$, $\alpha_{reverb}=6$) were empirically selected to balance speech intelligibility and environmental fidelity, and different operating points may be chosen depending on the target application.

When increasing $\alpha_{noise}$ with fixed $\alpha_{reverb}=6$, CLAP similarity first improved from 0.751 ($\alpha_{noise}=0$) to 0.770 ($\alpha_{noise}=3$) then decreased at higher values, while FAD remains relatively stable within the moderate range (2.36 to 2.90), indicating that the noise guidance primarily affects noise-related characteristics without substantially altering the reverberation reconstruction. 
However, excessive noise guidance leads to severe degradation in intelligibility, with WER rising sharply from 13.05 ($\alpha_{noise}=3$) to 31.53 ($\alpha_{noise}=6$) and 63.83 ($\alpha_{noise}=9$).

When increasing $\alpha_{reverb}$ with fixed $\alpha_{noise}=3$, FAD decreased dramatically from 10.38 ($\alpha_{reverb}=0$) to 1.48 ($\alpha_{reverb}=9$), while CLAP similarity remains relatively stable after $\alpha_{reverb}=3$ (0.758 to 0.770), confirming that the reverberation guidance primarily affects reverberation-related characteristics as captured by FAD, with limited impact on the noise component. 
This observation is consistent with the metric sensitivity analysis in Table~\ref{tab:metric_analysis}, where FAD was shown to be more sensitive to reverberation than CLAP similarity.

These results verify that the TCFG mechanism enables effective and independent control over the noise and reverberation components of the synthesized environmental speech.

\begin{table}[t]
\caption{Effect of TCFG scales ($\alpha_{speech}$:$\alpha_{noise}$:$\alpha_{reverb}$) on environment-aware generation. $\ast$ denotes the configuration used in the main experiments.}
\label{tab:tcfg}
\centering
\resizebox{\columnwidth}{!}{%
\begin{tabular}{lcccc}
\toprule
$\alpha_{s}$:$\alpha_{n}$:$\alpha_{r}$ & WER (\%) & SECS $\uparrow$ & CLAP SIM $\uparrow$ & FAD (VGGish) $\downarrow$ \\
\midrule
\multicolumn{5}{c}{\textit{Varying $\alpha_{noise}$ (fixed $\alpha_{speech}=1$, $\alpha_{reverb}=6$)}} \\
\midrule
1:0:6 & 8.32 & 0.387 & 0.751 & 2.36 \\
1:3:6$^\ast$ & 13.05 & 0.361 &  \textbf{0.770} & \textbf{2.11} \\
1:6:6 & 31.53 & 0.299 & 0.741 & 2.90 \\
1:9:6 & 63.83 & 0.209 & 0.674 & 4.43 \\
\midrule
\multicolumn{5}{c}{\textit{Varying $\alpha_{reverb}$ (fixed $\alpha_{speech}=1$, $\alpha_{noise}=3$)}} \\
\midrule
1:3:0 & 5.87 & 0.403 & 0.722 & 10.38 \\
1:3:3 & 7.35 & 0.386 & 0.758 & 4.18 \\
1:3:6$^\ast$ & 13.05 & 0.361 & 0.770 & 2.11 \\
1:3:9 & 25.64 & 0.334 & 0.769 & \textbf{1.48} \\
\bottomrule
\end{tabular}%
}
\end{table}
\section{Conclusions and Future Work}

This paper presented an extended DAIEN-TTS framework for environment-aware zero-shot TTS towards real-world scenarios.
 The framework jointly models speech, noise, and reverberation through an SES module and cross-attention-based environment conditioning mechanism, enabling independent control over the timbre and acoustic environment via separate audio prompts.
 A two-stage training pipeline, pretraining on simulated data and fine-tuning on real-world data, was adopted to bridge the simulated-to-real domain gap.

The experimental results indicate that joint modeling of noise and reverberation proved critical for both environment-robust and environment-aware synthesis, with DAIEN-TTS achieving a MOS comparable to the ground truth under simulated prompts and the highest environmental fidelity with an ESMOS approaching the ground-truth reference.
Fine-tuning on real-world data effectively improved both speaker similarity and environmental fidelity.
 The TCFG mechanism enabled largely independent control over the noise and reverberation components, and our analysis further revealed that CLAP similarity and FAD capture complementary aspects of environmental fidelity.

For future work, we plan to explore alternative environment conditioning modalities such as text descriptions of acoustic scenes and investigate the application of the DAIEN-TTS to synthetic training data generation for downstream tasks such as ASR and ASV.


%





\ifCLASSOPTIONcaptionsoff
  \newpage
\fi

\bibliographystyle{IEEEtran}
\bibliography{mybib}

\end{document}